\documentclass[aps,prd,showpacs]{revtex4}
\usepackage{amsfonts}
\usepackage{verbatim}
\usepackage{amsmath}
\usepackage[USenglish]{babel}
\usepackage[dvips]{epsfig}
\usepackage{graphicx}

\allowdisplaybreaks

\begin{document}

\begin{flushright}
{\tt FTPI-MINN-15/21} \\ 
\end{flushright}

\title{The full CPT-even photon sector of the Standard Model Extension at finite temperature \\
}
\author{Carlos A. Escobar$^{1}$ and Marcos A.~G. Garcia$^{2}$}
\affiliation{$^{1}$ Instituto de Ciencias Nucleares, Universidad Nacional Aut{\'o}noma de
M{\'e}xico, A. Postal 70-543, 04510 M{\'e}xico D.F., M{\'e}xico,}
\affiliation{$^{2}$ William I. Fine Theoretical Physics Institute, School of Physics and Astronomy, University of Minnesota, Minneapolis, MN 55455, USA}

\begin{abstract}
We study the finite temperature behavior of the CPT-even pure-photon sector of the Standard Model Extension, which is defined by the standard Maxwell Lagrangian plus the term $(k_F)_{\mu\nu\alpha\beta}F^{\mu\nu}F^{\alpha\beta}$. The Hamiltonian analysis is performed, from which the degrees of freedom and constraints of the theory are derived. We have explicitly calculated the partition function for an arbitrary configuration of the $(k_F)_{\mu\nu\alpha\beta}$ coefficients, to second order, and we have used it to obtain the thermodynamic properties of the modified photon sector. We find the correction to the frequency dependence in Planck's radiation law, and we identify that the total energy density is adjusted, re\-la\-tive to the standard scenario, by a global proportionality constant containing the Lorentz-violating contributions. Nevertheless, the equation of state is not affected by these modifications.
\end{abstract}

\maketitle

\section{Introduction}

One of the cornerstones of quantum field theories and general relativity is Lorentz invariance, which is assumed to be an exact symmetry. There is a lot of remarkably strong experimental support for this idea \cite{experiments}, with no violation detected. Nevertheless, certain quantum-gravity theories possess mechanisms that can lead to Lorentz violation \cite{quantum}, in which effects arising from modifications of space-time are expected to appear at distances of the order of the Planck length. This has attracted considerable attention in recent years both from the experimental and theoretical perspectives, given that, in principle, this would allow a better understanding about the space-time structure. The Standard Model Extension (SME) performed by Kostelecky et al.~\cite{SME} is a framework where Lorentz violation is motivated via spontaneous symmetry breaking (SSB), in which non-dynamical tensor fields are now added to General Relativity and the Standard Model and whose fixed directions induce the corresponding symmetry breaking in a given reference frame. These tensor fields are assumed to arise from non zero vacuum expectation values of some basic fields belonging to a more fundamental model, such as string theory \cite{String}.
 Considering the SME as a framework motivated from a SSB, the Goldstone theorem ensures that massless particles will emerge \cite{Goldstone}. Bjorken proposed that the photon can be a Goldstone mode associated with the SSB of Lorentz invariance \cite{Bjorken}. However, since the pure-photon sector of the SME is $U(1)$ gauge invariant, all particles are massless, and an alternative interpretation of the Goldstone theorem is required \cite{UrruEsc}. In fact, in the spontaneously broken space-time symmetries case, the counting of massless modes has to be done carefully \cite{BrokenSpace}, and it may happen that one to six Goldstone modes appear, each corresponding to one of the six Lorentz generators. The properties of Goldstone modes are, in general, model dependent and the knowledge of the fundamental theory is required to do a complete description. In the case of the SME some general conclusions about Goldstone modes have been obtained in \cite{Bluhm1}.

The pure-photon sector of the SME includes the usual Maxwell term plus the presence of the CPT-odd term ($\frac{1}{2}(k_{AF})^\kappa\epsilon_{\kappa\lambda\mu\nu}A^\lambda F^{\mu\nu} $), sometimes called the Carroll-Field-Jackiw term \cite{Carrollterm}, and the CPT-even term $-\frac{1}{4}(k_F)_{\mu\nu\alpha\beta}F^{\mu\nu}F^{\alpha\beta}$. Both terms have been extensively studied in the literature \cite{Studyphot1}, and experimental constraints exist for them \cite{testphotonsector}. The search for new effects arising from these Lorentz violating terms, and an improvement of the bounds for the magnitude of these coefficients constitute two of the main lines of study. The study of the cosmic microwave background (CMB) offers an opportunity to test the pure-photon sector of the SME at finite temperature \cite{SMETempFin}, since the propagation of light would be affected, in the form of non-standard dispersion relations, polarization, birefringence properties, among other effects \cite{SME,Studyphot1,Studyphot2,testphotonsector}. As it may be expected, thermodynamic properties and the spectral distribution can be modified as well. In Ref. \cite{CFJTerm}, the partition function in the functional integral formalism was calculated in order to study the finite temperature behavior of the Carroll-Field-Jackiw term, for the case of a purely spacelike background. In Ref. \cite{Termeven} the study was extend to the CPT-even term for particular configurations of the coefficients $(k_F)_{\mu\nu\alpha\beta}$, to simplify the calculations. In both cases, Lorentz violation corrections to the blackbody radiation and anisotropy in the angular distribution for the energy density were found. It remains unclear, however, if there is no information loss from the consideration of only a few particular configurations. The goal of this paper is to study the finite temperature properties of the CPT-even pure-photon sector of the SME in the most general case, following a scheme similar to that employed in \cite{CFJTerm,Termeven}. The outline of this work is the following. In Section \ref{Modelo} the CPT-even pure-photon sector of the SME Lagrangian is introduced and some properties are reviewed. Following Dirac's scheme for constrained systems, the Hamiltonian analysis is performed, and the canonical quantization is carried out. In Section \ref{FunctionPart} the partition function is evaluated in the functional formalism and some thermodynamical properties are derived, a substantial improvement over the previous calculations is reported considering an arbitrary configuration for the $(k_F)_{\mu\nu\alpha\beta}$ coefficients. We compare the result with a classical thermodynamic approach. Our summary and conclusions are contained in Sec. \ref{summary}.

\section{The Model}\label{Modelo}

We focus on the pure-photon sector, and particularly on the CPT-even violating terms within the minimal SME. The Lagrangian density is given by
\begin{equation}
L=-\frac{1}{4}F_{\mu\nu}F^{\mu\nu}-\frac{1}{4}(k_F)_{\mu\nu\alpha\beta}F^{\mu\nu}F^{\alpha\beta}\, ,
\label{lagrangiano}
\end{equation}
where the first term corresponds to the standard electrodynamics, being $F_{\mu\nu}=\partial_\mu A_\nu-\partial_\nu A_\mu$ the electromagnetic stress tensor. The second term introduces Lorentz-CPT symmetry breakdown, which is controlled by the non-dynamical spacetime-constant and dimensionless quantities $(k_F)_{\mu\nu\alpha\beta}$; these coefficients have the symmetries of the Riemann tensor and a vanishing double trace, which imply a total of 19 independent components. The tensor $(k_F)_{\mu\nu\alpha\beta}$ is alternatively parametrized in terms of four $3\times3$ matrices \cite{Studyphot1}, $\kappa_{DE},\kappa_{HB},\kappa_{DB},\kappa_{HE}$, defined by
\begin{equation}
(\kappa_{DE})^{jk}=-2(k_F)^{0j0k},\quad(\kappa_{HB})^{jk}=\frac{1}{2}\epsilon^{jpq}\epsilon^{klm}(k_F)^{pqlm},\quad
(\kappa_{DB})^{jk}=-(\kappa_{HE})^{kj}=\epsilon^{kpq}(k_F)^{0jpq}.
\label{parametrization1}
\end{equation}
The matrices $\kappa_{DE}$ and $\kappa_{HB}$ contain together $11$ independent components, while $\kappa_{DB}$ and $\kappa_{HE}$ possess together $8$ components, which encompass the $19$ independent elements of the tensor $(k_F)_{\mu\nu\alpha\beta}$. An alternative parametrization, which allows easier experimental constraints, consists of writing $(k_F)_{\mu\nu\alpha\beta}$ in terms of four traceless matrices and one trace element \cite{Studyphot1},
\begin{equation}
(\tilde{\kappa}_{o+})^{jk}=\frac{1}{2}(\kappa_{DB}+\kappa_{HE})^{jk},\quad\quad
(\tilde{\kappa}_{o-})^{jk}=\frac{1}{2}(\kappa_{DB}-\kappa_{HE})^{jk},
\label{parametrization21}
\end{equation}
\begin{equation}
(\tilde{\kappa}_{e+})^{jk}=\frac{1}{2}(\kappa_{DE}+\kappa_{HB})^{jk},\quad\quad
(\tilde{\kappa}_{e-})^{jk}=\frac{1}{2}(\kappa_{DE}-\kappa_{HB})^{jk}-\frac{1}{3}\delta^{jk}\textrm{Tr}(\kappa_{DE}),\quad\quad \tilde{\kappa}_{\textrm{tr}}=\frac{1}{3}\textrm{Tr}(\kappa_{DE}).
\label{parametrization22}
\end{equation}
All parity-even coefficients are contained in $\tilde{\kappa}_{e+},\tilde{\kappa}_{e-}$ and $\tilde{\kappa}_{\textrm{tr}}$, while all parity-odd coefficients are contained in $\tilde{\kappa}_{o+}$ and $\tilde{\kappa}_{o-}$. The matrix $\tilde{\kappa}_{o+}$ is antisymmetric while the remaining matrices are symmetric. In Section \ref{FunctionPart} we will use this second parametrization to express our main results.

As we previously mentioned, the $(k_F)_{\mu\nu\alpha\beta}$ coefficients can be motivated by spontaneous breaking of Lorentz symmetry \cite{String}, avoiding the issues of incompatibility in General Relativity present when an explicit Lorentz symmetry violation is introduced \cite{Noexplicit}. The transformation $A_\mu\rightarrow A_\mu+\partial_\mu \Lambda$ leaves the Lagrangian (\ref{lagrangiano}) invariant, and therefore, the gauge symmetry $U(1)$ is preserved as in the Maxwell theory. We use the convention, Greek indices  $\mu,\nu=0,1,2,3$, Latin indices $i,j=1,2,3$ and the metric $\eta_{\mu\nu}=(1,-1,-1,-1)$. The Euler-Lagrange equations arising from the Lagrangian (\ref{lagrangiano}) correspond to
\begin{equation}
\partial^\nu F_{\mu\nu}+ (k_F)_{\mu\nu\alpha\beta} \partial^\nu F^{\alpha\beta}=0\, .
\label{eqLagrange}
\end{equation}
The remaining Maxwell equations
\begin{equation}
\partial_\mu \tilde{F}^{\mu\nu} \equiv \frac{1}{2} \epsilon^{\mu\nu\alpha\beta}\partial_\mu F_{\alpha\beta}=0,
\end{equation}
continue to hold as a consequence of defining $F^{\mu\nu}$ through the potential $A_\mu$. As we previously stated, the propagation of light is modified due the presence of the $(k_F)_{\mu\nu\alpha\beta}$ coefficients, in this case the dispersion relation for the Lagrangian (\ref{lagrangiano}) is given by \cite{Studyphot1}
\begin{equation}
p^0_{\pm}=(1+\rho\pm\sigma)|\vec{p}|,\quad\quad
\rho=-\frac{1}{2}\tilde{k}_\alpha\,^\alpha,\quad\quad\sigma^2=\frac{1}{2}(\tilde{k}_{\alpha\beta})^2-\rho^2,
\label{dispersion}
\end{equation}
to lowest order in $(k_F)_{\mu\nu\alpha\beta}$, with
\begin{equation}
\tilde{k}^{\alpha\beta}=(k_F)^{\alpha\mu\beta\nu}\hat{p}_\mu \hat{p}_\nu,\quad\quad \hat{p}^\mu=\frac{p^\mu}{|\vec{p}|}\, .
\end{equation}

Let us now consider the analysis of constraints {\em \`a la} Dirac \cite{Dirac}, and the canonical quantization of the model. This will allow us to determine the number of degrees of freedom and establish some differences between the standard electrodynamics and the CPT-even pure-photon sector of the SME. The canonically conjugated momenta are given by
\begin{eqnarray}
\label{momenta}
\pi^i\equiv\frac{\partial L}{\partial \dot{A}_i} &=& F^{i0}+(k_F)^{i0\alpha\beta}F_{\alpha\beta} \\ \nonumber
&=& F^{j0}(\delta^i\,_j+2(k_F)^{i0}\,_{j0}) +(k_F)^{i0lm}F_{lm} \\ \nonumber
&=& M^i\,_j F^{j0}+N^i\, ,
\end{eqnarray}
where we have defined $M^i\,_j\equiv\delta^i\,_j+2(k_F)^{i0}\,_{j0}$ and $N^i\equiv(k_F)^{i0lm}F_{lm}$. The canonical momentum associated to $A_0$ is null, $\pi^0=0$, as it is in standard electrodynamics. Approximating all quantities to first order in the Lorentz-violating coefficients, we find that the matrix $M^i\,_j$ has the inverse $B^i\,_j\equiv(M^{-1})^i\,_j=\delta^i\,_j-2(k_F)^{i0}\,_{j0}$, which allows us to rewrite (\ref{momenta}) as $F^{k0}=B^{k}\,_i \pi^i-N^k$. Using the above it is straightforward to obtain the canonical Hamiltonian density
\begin{equation}
H_c=-\frac{1}{2}B^i\,_j \pi^j\pi_i+\frac{1}{4}F_{ij}F^{ij}-(k_F)^{0ilm}\pi_iF_{lm}+\frac{1}{4}(k_F)^{ijlm}F_{ij}F_{lm}-A_0\partial^k\pi_k\, ,
\label{hamiltonian_can}
\end{equation}
where we have carried out one integration by parts and omitted boundary terms. The non-zero Poisson brackets (PB) are given by
\begin{equation}
\{A_\mu(\mathbf{x},t),\pi^\nu(\mathbf{y},t)\}_P=\delta_\mu\,^\nu\delta^3(\mathbf{x}-\mathbf{y})\, .
\end{equation}

In what follows we will assume that all PB are calculated at equal times and we omit the label $t$. We employ Dirac's method to construct the canonical theory due to the
fact that the primary constraint
\begin{equation}
\phi_1=\pi ^{0}\simeq 0,
\label{pizero}
\end{equation}
is present (here the symbol $\simeq$ denotes the weak equality). The extended Hamiltonian density is defined as
\begin{equation}
H_E=H_c+\lambda\phi_1,
\end{equation}
where $\lambda$ is an arbitrary function. The evolution condition of the
primary constraint (\ref{pizero}),
\begin{equation}
\dot{\phi_1}(\mathbf{x})=\{\phi_1(\mathbf{x}),\int d^{3}y\;H_{E}(\mathbf{y})\}_P\simeq 0\, ,
\end{equation}
leads to Gauss' law,
\begin{equation}
\phi_2=\partial_i\pi^i\simeq 0\, .
\label{gausslaw}
\end{equation}
It is not difficult to prove that\ (\ref{pizero}) and (\ref{gausslaw}) are the
only constraints present in the model, and that they constitute a first class set ($\{\phi_1,\phi_2\}=0$). As in standard electrodynamics, the model possesses two degrees of freedom (DOF), following Dirac's scheme, $\textrm{DOF}=\frac{1}{2}[\textrm{variables in the phase space}-\textrm{second class constrictions}-2\times\textrm{first class constrictions}]=\frac{1}{2}[8-0-2\times2]=2$. If we write (\ref{gausslaw}) in terms of the field strength using the definition of the canonical momenta (\ref{momenta}), we obtain
\begin{equation}
\partial_i\pi^i=\partial_iF^{i0}+(k_F)^{i0\alpha\beta}\partial_iF_{\alpha\beta}=0\,.
\end{equation}
We recognize the last equation as the Lagrangian equation (\ref{eqLagrange}) for the $\mu=0$ component; however, in the Hamiltonian formalism it is a constraint, not an equation of motion. Gauss' law (\ref{gausslaw}) is different from its standard electrodynamics form, being this even more evident when it is rewritten in terms of the electric and magnetic fields instead of the canonical momenta $(\pi^i=F^{i0}+(k_F)^{i0\alpha\beta}F_{\alpha\beta})$. In order to construct a quantum theory via canonical quantization ($\{A,B\}\rightarrow \frac{1}{i\hbar}[\hat{A},\hat{B}]$), we must remove the extra degrees of freedom. This means that we have to impose as many suitable gauge constraints ``by hand" as there are first class constraints; these gauge constraints have to be admissible and convert the first class constraints into second class constraints, and then we can introduce the Dirac brackets to perform the correct quantization. We choose the Coulomb gauge ($\Phi_1=\partial_iA^i\simeq0$) plus $\Phi_2=A_0\simeq0$ to fix the gauge. The Dirac brackets
\begin{equation}
\{A(\textbf{x}),B(\textbf{y})\}_D=\{A(\textbf{x}),B(\textbf{y})\}_P - \int\,d^3u\,d^3v \{A(\textbf{x}),\chi_i(\textbf{u})\}_P(Q^{-1})^{ij}\{\chi_j(\textbf{v}),B(\textbf{y})\}_P\,,
\end{equation}
where $\chi_i$ is one of the constraints ($\phi_1,\phi_2,\Phi_1,\Phi_2$) and $Q^{ij}(\mathbf{x},\mathbf{y})=\{\chi_i(\mathbf{x}),\chi_j(\mathbf{y})\}_P$,
\begin{equation}
Q^{ij}(\mathbf{x},\mathbf{y})= \left(
\begin{array}{cccc}
0 & 0 & -1 & 0 \\
0 & 0 & 0 & \nabla^2 \\
1 & 0 & 0 & 0 \\
0 & -\nabla^2 & 0 & 0 \\
\end{array}
\right) \delta^3(\mathbf{x}-\mathbf{y});\quad\quad
(Q^{-1})^{ij}(\mathbf{x},\mathbf{y})= \left(
\begin{array}{cccc}
0 & 0 & \delta^3(\mathbf{x}-\mathbf{y}) & 0 \\
0 & 0 & 0 & \frac{1}{4\pi|\mathbf{x}-\mathbf{y}|} \\
-\delta^3(\mathbf{x}-\mathbf{y}) & 0 & 0 & 0 \\
0 & -\frac{1}{4\pi|\mathbf{x}-\mathbf{y}|} & 0 & 0 \\
\end{array}
\right),
\end{equation}
are given by
\begin{eqnarray}
\label{BracketD}
&&\{A_i(\mathbf{x},t),A^j(\mathbf{y},t)\}_D=0, \\ \nonumber
&&\{\pi_i(\mathbf{x},t),\pi^j(\mathbf{y},t)\}_D=0, \\ \nonumber
&&\{A_i(\mathbf{x},t),\pi^j(\mathbf{y},t)\}_D=\delta_i\,^j\delta(\mathbf{x}-\mathbf{y})
+\partial_i\partial^j\frac{1}{4\pi|\mathbf{x}-\mathbf{y}|}\equiv \delta^3_{\perp i}\,^j (\mathbf{x}-\mathbf{y}),
\end{eqnarray}
and they have the same form as in standard electrodynamics when the canonical momenta are used. However, rewriting in terms of the electric and magnetic fields, the difference is manifest,
\begin{eqnarray}
&&\{A_i(\mathbf{x},t),A^j(\mathbf{y},t)\}_D=0, \\ \nonumber
&&\{E_i(\mathbf{x},t),E^j(\mathbf{y},t)\}_D=2[(k_F)^{i0lj}\partial_{xl}-
(k_F)^{j0li}\partial_{yl}]\delta^3(\mathbf{x}-\mathbf{y}), \\ \nonumber
&&\{A_i(\mathbf{x},t),E^j(\mathbf{y},t)\}_D=-\delta^3_{\perp i}\,^j (\mathbf{x}-\mathbf{y})+2(k_F)^{j0}\,_{s0}\delta^3_{\perp i}\,^s (\mathbf{x}-\mathbf{y}).
\end{eqnarray}
Once the Dirac brackets have been included, the dynamics of the theory will be generated by the Hamiltonian (\ref{hamiltonian_can}) without the term $A_0\partial_i\pi^i$, which is proportional to $\phi_2$ and has already been fixed and included in the process of introducing the Dirac brackets. The canonical quantization can be carried out using the aforementioned Hamiltonian (\ref{hamiltonian_can}) and the brackets given by (\ref{BracketD}). Being $U(1)$ the group of symmetry of the theory, there are other possibilities to fix the gauge, as the Lorentz gauge ($\partial_\mu A^\mu$=0), which is manifestly covariant; however, it is not possible to handle such a gauge with the quantization formalism that we employ here. This is due to the fact that the Lorentz gauge involves the time derivative of $A_0$. There are well known formalisms which are capable of dealing with such relativistic constraints, among which are BRST quantization \cite{BRST}, the Fadeev-Popov method \cite{Faddeev} into the path integral \cite{FieldTheory} or within the Hamiltonian formalism, one has \cite{GaugeCov}. Nevertheless, these approaches are not within of the scope of the present work.

\bigskip

Rewriting the Hamiltonian (\ref{hamiltonian_can}) in terms of the electric and magnetic fields we find
\begin{equation}\label{HEB}
H= \frac{1}{2}(\mathbb{E}^2+\mathbb{B}^2) - (k_F)^{0j0k}E^jE^k+\frac{1}{4}(k_F)^{jklm}\epsilon^{jkp}\epsilon^{lmq}B^pB^q\,,
\end{equation}
where $\mathbb{E}^2=E_iE_i$, $B_i=-\frac{1}{2}\epsilon^{ijk}F^{jk}$, and therefore $\mathbb{B}^2=\frac{1}{2}F_{ij}F^{ij}$. The same result has been found in \cite{SME} following a different line of thought, where it was shown that if $(k_F)_{\mu\nu\alpha\beta}$ is small, the last quantity (\ref{HEB}) is nonnegative. This is due to the fact that the Hamiltonian (\ref{HEB}) can be viewed as the bilinear form $x^TMx$ with $x^T\equiv(\mathbb{E},\mathbb{B})$. It can be readily shown that, upon diagonalization, the matrix $M$ has entries $\frac{1}{2}-O(k_F)>0$ for both the electric and magnetic field contributions.

\section{Partition Function and Thermodynamics}\label{FunctionPart}

We derive now some of the thermodynamic properties of the Lagrangian (\ref{lagrangiano}). Our main goal is to obtain the finite temperature energy density of the electromagnetic field. Following the quantum field theory scheme, we calculate the partition function. In the previous section we adopted the Coulomb gauge; hereafter we will switch to a covariant gauge. The simplest way to obtain the partition function is trough the Faddeev-Popov method \cite{Faddeev}, which is equivalent to the introduction of constraints as done in Sec.~\ref{Modelo}; both methods allow us to work with the effective degrees of freedom. Choosing the Lorentz gauge we can write the partition function as
\begin{equation}
\textrm{Z}=\int [dA^\mu]\,\textrm{det}(-\partial^2)\,\textrm{exp}\bigg(\int_0^\beta d\tau\int d^3x \,L_{\rm eff}\bigg),
\end{equation}
where $\textrm{det}(-\partial^2)$ is the Faddeev-Popov determinant and we have switched to an imaginary time variable $\tau=it$. The effective Lagrangian is given by
\begin{eqnarray}
L_{\rm eff}&=&-\frac{1}{4}F_{\mu\nu}F^{\mu\nu}-\frac{1}{4}(k_F)_{\mu\nu\alpha\beta}F^{\mu\nu}F^{\alpha\beta}-\frac{1}{2\rho}
(\partial_\mu A^\mu)^2\, .
\end{eqnarray}
Upon substitution of the finite temperature replacements $t\rightarrow-i\tau$, $A_0\rightarrow iA_\tau$ and $(k_F)^{0\mu\nu\alpha}\rightarrow i(k_F)^{\tau\mu\nu\alpha}$ (similar convention to other indices in $(k_F)^{\mu\nu\alpha\beta}$), the effective Lagrangian can be written in Euclidean notation, with $\mu,\nu,\alpha,\beta=\tau,1,2,3$,
\begin{eqnarray}
L_{\rm eff}&=& \frac{1}{2}A_\nu\bigg[\delta_{\mu\nu}\partial^2-\bigg(1-\frac{1}{\rho}\bigg)\partial_\mu\partial_\nu+
2(k_F)_{\beta\nu\alpha\mu}\partial_\beta\partial_\alpha \bigg]A_\mu\quad\quad\quad (\rho\rightarrow1)
\\ \nonumber  &=& \frac{1}{2}A_\nu\bigg[\delta_{\mu\nu}\partial^2+
2(k_F)_{\beta\nu\alpha\mu}\partial_\beta\partial_\alpha \bigg]A_\mu \\ \nonumber
&=&\frac{1}{2}\textrm{A}^{T}\textrm{DA}\,.
\end{eqnarray}
In the first line we have chosen the Feynman gauge ($\rho=1$) and $D_{\mu\nu}=\delta_{\mu\nu}\partial^2+
2(k_F)_{\beta\nu\alpha\mu}\partial_\beta\partial_\alpha$. The field admits a Fourier expansion:
\begin{equation}
A_\mu(\tau,\textbf{x})=\sqrt{\frac{\beta}{V}}\sum_{n,p}e^{i(\omega_n\tau+\textbf{x}\cdot\textbf{p})} \tilde{A}_\mu(n,\textbf{p}),
\label{Fourier}
\end{equation}
where $\omega_n=\frac{2\pi n}{\beta}$ are the Matsubara frequencies, and the field $A_\mu(\tau,\textbf{x})$ satisfies the constraints of periodicity $A_\mu(\beta,\textbf{x})=A_\mu(0,\textbf{x})$ for all $\textbf{x}$. The normalization in (\ref{Fourier}) is chosen so that each Fourier amplitude is dimensionless. If we use a ghost field $C$ to write
\begin{equation}
\textrm{det}(-\partial^2)=\int[d\bar{C}][dC]\textrm{exp}\bigg(\int d\tau\int d^3x(\partial_\mu \bar{C})(\partial^ \mu C)\bigg),
\end{equation}
then we can calculate the partition function in frequency-momentum space as
\begin{equation}
\textrm{ln Z}=\textrm{Tr}\,\textrm{ln}[\beta^2(\omega^2_n+\mathbf{p}^2)]-\frac{1}{2}\textrm{ln}[\textrm{Det(D)}],
\end{equation}
where now
\begin{equation}
D= \beta^2\left(
\begin{array}{cccc}
\omega^2_n+\mathbf{p}^2+\Lambda_{\tau\tau} & \Lambda_{\tau x} & \Lambda_{\tau y} & \Lambda_{\tau z} \\
\Lambda_{\tau x} & \omega^2_n+\mathbf{p}^2+\Lambda_{xx} & \Lambda_{x y} &\Lambda_{x z} \\
\Lambda_{\tau y} & \Lambda_{x y} & \omega^2_n+\mathbf{p}^2+\Lambda_{yy} & \Lambda_{yz} \\
\Lambda_{\tau z} & \Lambda_{x z} & \Lambda_{yz} & \omega^2_n+\mathbf{p}^2+\Lambda_{zz} \\
\end{array}
\right),
\end{equation}
and $\Lambda_{\mu\nu}=2(k_F)_{\mu\alpha\nu\beta} p_\alpha p_\beta$, ($p_\tau\equiv\omega_n$). Calculation of the determinant to second order in $k_F$ gives
\begin{eqnarray}
\textrm{det}(D)&=&\prod_{n,\textbf{p}}\beta^8[(\omega^2_n+\textbf{p}^2)^4]\bigg(1
+\textrm{Tr}(\tilde{\Lambda})+\frac{1}{2}(\textrm{Tr}(\tilde{\Lambda}))^2-\frac{1}{2}\textrm{Tr}(\tilde{\Lambda}^2)\bigg)\,,
\end{eqnarray}
where we have defined $\tilde{\Lambda}_{\mu\nu}=2(k_F)_{\mu\alpha\nu\beta} p_\alpha p_\beta/(\omega^2_n+\textbf{p}^2)$ and the relation
 \begin{equation}
 \textrm{Det}(1+M)= 1+\textrm{Tr}(M)+\frac{1}{2} (\textrm{Tr}(M))^2-\frac{1}{2}\textrm{Tr}(M^2)+O(M^3), 
\end{equation} 
has been employed. Therefore, the total partition function becomes
\begin{eqnarray}
\label{paritionF2}
\textrm{ln Z}&=&-\sum_{n,\mathbf{p}}\textrm{ ln}[\beta^2(\omega^2_n+\mathbf{p}^2)]-\frac{1}{2}\sum_{n,\mathbf{p}}\textrm{ ln}\bigg(1
+\textrm{Tr}(\tilde{\Lambda})+\frac{1}{2}(\textrm{Tr}(\tilde{\Lambda}))^2-\frac{1}{2}\textrm{Tr}(\tilde{\Lambda}^2)+\cdots\bigg)\, \\ \nonumber
&\equiv& \textrm{ln Z}_0 + \textrm{ln Z}_{LV}\, .
\end{eqnarray}
We recognize the first term as the usual result for the Maxwell theory, which corresponds to massless bosons with two spin degrees of freedom in thermal equilibrium; in other words, blackbody radiation. All modifications to the standard case due to Lorentz violation come from to the second term in (\ref{paritionF2}), which we can evaluate as follows
\begin{equation}
\textrm{ln Z}_{LV}=-\frac{1}{2}\sum_{n,\mathbf{p}}\textrm{ ln}\bigg(1
+\textrm{Tr}(\tilde{\Lambda})+\frac{1}{2}(\textrm{Tr}(\tilde{\Lambda}))^2-\frac{1}{2}\textrm{Tr}(\tilde{\Lambda}^2)\bigg) \approx -\frac{1}{2}\sum_{n,\mathbf{p}}\bigg(
\textrm{Tr}(\tilde{\Lambda})-\frac{1}{2}\textrm{Tr}(\tilde{\Lambda}^2)\bigg) 
\equiv-\sum_{n,\textbf{p}} (\bar{Z}_{LV_1}+\bar{Z}_{LV_2}).
\end{equation}
Here we have defined $\bar{Z}_{LV_1}$ and $\bar{Z}_{LV_2}$ as the Lorentz violation contributions to first and second order, respectively. We begin calculating the first order contributions as follows:
\begin{eqnarray}
-\sum_{n,\textbf{p}} \bar{Z}_{LV_1}&=& -\frac{1}{2}\sum_{n,\textbf{p}}\textrm{Tr}(\tilde{\Lambda}) \nonumber \\ \nonumber &=&-\sum_{n,\textbf{p}}\frac{(k_F)_{\alpha\mu\alpha\nu}p_\mu p_\nu}{(\omega^2_n+\textbf{p}^2)}\quad\quad\quad\quad\quad(p_0=\omega_n=\frac{2\pi n}{\beta}) \\ \nonumber
&=& -\sum_{n,\mathbf{p}} \frac{1}{(\omega^2_n+\textbf{p}^2)}\bigg[(k_F)_{\alpha \tau\alpha\tau}\omega_n^2+2(k_F)_{\alpha i\alpha \tau}\omega_np_i+(k_F)_{\alpha i\alpha j}p_i p_j\bigg] \\
&=& -\sum_{n,\mathbf{p}} \frac{1}{(\omega^2_n+\textbf{p}^2)}\bigg[(k_F)_{\alpha \tau\alpha \tau}\omega_n^2+(k_F)_{\alpha i\alpha j}p_i p_j\bigg]\,. \label{cxh1}
\end{eqnarray}
where in the third line we employed the sum $\sum_{n=-\infty}^\infty\frac{n}{n^2+a^2}=0$. Adding and subtracting the term $(k_F)_{\alpha \tau\alpha \tau}\mathbf{p}^2$ inside the brackets in (\ref{cxh1}), we arrive to the equivalent expression
\begin{eqnarray}
-\sum_{n,\textbf{p}} \bar{Z}_{LV_1}
&=& -\sum_{n,\mathbf{p}} \bigg[(k_F)_{\alpha \tau\alpha \tau}+\frac{-(k_F)_{\alpha \tau\alpha \tau}\mathbf{p}^2+(k_F)_{\alpha i\alpha j}\hat{p}_i \hat{p}_j\mathbf{p}^2}{(\omega^2_n+\textbf{p}^2)}\bigg], \\ \nonumber
\end{eqnarray}
with $\hat{p}_i=p_i/|\vec{p}|$. Making use of the identity
\begin{equation}
\sum_{n=-\infty}^{\infty}\frac{1}{\omega^2_n+\mathbf{p}^2}=\frac{\beta}{2\mathbf{p}}\textrm{coth}\bigg(\frac{\beta \mathbf{p}}{2}\bigg) =\frac{\beta}{2\mathbf{p}}\bigg(1+\frac{2}{e^{\beta \mathbf{p}}-1}\bigg),
\end{equation}
we find that the contribution due to Lorentz violation to first order can be written as
\begin{equation}\label{zlv}
-\sum_{n,\textbf{p}} \bar{Z}_{LV_1}=-V\int \frac{d^3\textrm{p}}{(2\pi)^3}\, [-(k_F)_{\alpha \tau\alpha \tau}+(k_F)_{\alpha i\alpha j}\hat{p}_i \hat{p}_j]\mathbf{p}^2\bigg(\frac{\beta}{2\mathbf{p}}\bigg)\bigg(1+\frac{2}{e^{\beta \mathbf{p}}-1}\bigg)\,,
\end{equation}
where a temperature-independent divergent term has been dropped; it is well known that any quantity in finite temperature theory is defined after subtraction by its $T=0$ counterpart, and so $T$-independent parts of the partition function, infinite or finite, are of no importance \cite{Tinfinity} . Taking the standard spherical coordinate system and $|\mathbf{p}|=\omega$, eq. (\ref{zlv}) becomes
\begin{equation}\label{sphc}
-\sum_{n,\textbf{p}} \bar{Z}_{LV_1}=-\frac{V}{(2\pi)^3}\int d\omega\, d\Omega\, [-(k_F)_{\alpha \tau\alpha \tau}+(k_F)_{\alpha i\alpha j}\hat{p}_i \hat{p}_j]\bigg(\frac{\beta\omega^3}{2}\bigg)\bigg(1+\frac{2}{e^{\beta \omega}-1}\bigg)\,,
\end{equation}
where $\hat{p}_1=\sin(\theta)\cos(\phi),\,\hat{p}_2=\sin(\theta)\sin(\phi)$ and $\hat{p}_3=\cos(\theta)$. Performing the integral over solid angle we find
\begin{equation}
-\sum_{n,\textbf{p}} \bar{Z}_{LV_1}=-4\pi \frac{V}{(2\pi)^3} [-(k_F)_{\alpha \tau\alpha \tau}+\frac{1}{3}(k_F)_{\alpha i\alpha i}]\int d\omega \, \bigg(\frac{\beta\omega^3}{2}\bigg)\bigg(1+\frac{2}{e^{\beta \omega}-1}\bigg).
\label{partinter}
\end{equation}
We now make use of the vanishing of the double trace of $(k_F)_{\mu\nu\alpha\beta}$ condition, which in Euclidean space implies
\begin{equation}
(k_F)^{\mu\nu}\,_{\mu\nu}=2(k_F)^{0i}\,_{0i}+(k_F)^{ij}\,_{ij}=-2(k_F)_{0i0i}+(k_F)_{ijij}\quad \xrightarrow{\quad-i\tau\quad}  \quad 2(k_F)_{\tau i\tau i}+(k_F)_{ijij}=0\, .
\label{vanishingtrace}
\end{equation}
Using the above, (\ref{partinter}) becomes
\begin{equation}
\label{nonstandardplanck}
-\sum_{n,\textbf{p}} \bar{Z}_{LV_1}=\frac{16\pi}{3}\frac{V}{(2\pi)^3}(k_F)_{\alpha \tau\alpha \tau}\int d\omega \, \bigg(\frac{\beta\omega^3}{2}\bigg)\bigg(1+\frac{2}{e^{\beta \omega}-1}\bigg) = 2\frac{\pi^2V}{45\beta^3}(k_F)_{\alpha \tau\alpha \tau} =3\frac{\pi^2V}{45\beta^3}\tilde{\kappa}_{\textrm{tr}},
\end{equation}
where we have neglected vacuum contributions to perform the integral. The above implies that the modification to the energy density due Lorentz violation to first order will have the same dependence in the temperature as the standard theory, $U\sim T^4$. The second order contribution can be readily evaluated in a similar way, 

\begin{equation}
-\sum_{n,\textbf{p}}\bar{Z}_{LV_2}= \frac{1}{4}\sum_{n,\textbf{p}} \textrm{Tr}(\tilde{\Lambda}^2)
=\sum_{n,\textbf{p}} A_1 \frac{\omega_n^4}{(\omega^2_n+\mathbf{p}^2)^2}+A_2 \frac{\textbf{p}^2\omega_n^2}{(\omega^2_n+\mathbf{p}^2)^2}+A_3\frac{\textbf{p}^4}{(\omega^2_n+\mathbf{p}^2)^2},
\end{equation}
where the momentum-dependent coefficients $A_{1,2,3}$ are defined as
\begin{equation}
\begin{aligned}
A_1&=(k_F)_{\mu\tau\nu\tau}(k_F)_{\mu\tau\nu\tau}\,, \\ 
A_2&=\Big( 2(k_F)_{\mu\tau\nu\tau}(k_F)_{\mu i\nu j}+[(k_F)_{\mu\tau\nu i}+(k_F)_{\mu i\nu\tau}][(k_F)_{\mu\tau\nu j}+(k_F)_{\mu j\nu\tau}]\Big) \hat{p}_i\hat{p}_j\,,\\
A_3&= (k_F)_{\mu i \nu j}(k_F)_{\mu l \nu m}\hat{p}_i\hat{p}_j\hat{p}_l\hat{p}_m\,.
\end{aligned}
\end{equation}
Making use of the identities
\begin{equation}
\sum_{n=-\infty}^{\infty}\frac{1}{(\omega_n^2+\textbf{p}^2)^2}=\frac{\beta^2}{8\textbf{p}^2}\textrm{csch}^2\bigg(\frac{|\textbf{p}|\beta}{2}\bigg)
+\frac{\beta}{4|\textbf{p}|^3}\bigg(1+\frac{2}{e^{\beta|\textbf{p}|}-1}\bigg),
\label{suma1}
\end{equation}
\begin{equation}
\sum_{n=-\infty}^{\infty}\frac{\omega_n^2}{(\omega_n^2+\textbf{p}^2)^2}=-\frac{\beta^2}{8}\textrm{csch}^2\bigg(\frac{|\textbf{p}|\beta}{2}\bigg)
+\frac{\beta}{4|\textbf{p}|}\bigg(1+\frac{2}{e^{\beta|\textbf{p}|}-1}\bigg),
\label{suma2}
\end{equation}
\begin{equation}
\sum_{n=-\infty}^{\infty}\frac{\omega_n^4}{(\omega_n^2+\textbf{p}^2)^2}=\sum_n\bigg(1
-\frac{2\omega_n^2\textbf{p}^2}{(\omega_n^2+\textbf{p}^2)^2}
-\frac{\textbf{p}^4}{(\omega_n^2+\textbf{p}^2)^2}\bigg),
\label{suma3}
\end{equation}
we now evaluate the sums over $n$. Employing the same spherical coordinate system as in (\ref{sphc}) we find
\begin{eqnarray} \nonumber
-\sum_{n,\textbf{p}}\bar{Z}_{LV_2}&=&V\int \frac{d^3p}{(2\pi)^3}\,\left[ (A_1-A_2+A_3)\frac{\textbf{p}^2\beta^2}{8}\textrm{csch}^2\bigg(\frac{|\textbf{p}|\beta}{2}\bigg)
+(-3A_1+A_2+A_3)\frac{|\textbf{p}|\beta}{4}\bigg(1+\frac{2}{e^{\beta|\textbf{p}|}-1}\bigg)\right] \\ \nonumber
&=&V\int \frac{d\Omega d\omega}{(2\pi)^3}\,\left[ (A_1-A_2+A_3)\frac{\omega^4\beta^2}{8} \textrm{csch}^2\bigg(\frac{\omega\beta}{2}\bigg)
+(-3A_1+A_2+A_3)\frac{\omega^3\beta}{4}\bigg(1+\frac{2}{e^{\beta\omega}-1}\bigg)\right]\\
&=&V\int \frac{ d\omega}{2\pi^2}\,\left[(\tilde{A}_1-\tilde{A}_2+\tilde{A}_3)\frac{\omega^4\beta^2}{8}\textrm{csch}^2\bigg(\frac{\omega\beta}{2}\bigg)
+(-3\tilde{A}_1+\tilde{A}_2+\tilde{A}_3)\frac{\omega^3\beta}{4}\bigg(1+\frac{2}{e^{\beta\omega}-1}\bigg)\right],
\label{2contribution}
\end{eqnarray}
where now the momentum-independent coefficients $\tilde{A}$ correspond to
\begin{align}
\tilde{A}_1 &=  A_1 =\frac{1}{4}\textrm{Tr}(\kappa_{DE}^2), \\ \nonumber
\tilde{A}_2 &= \frac{2}{3}\bigg((k_F)_{\mu \tau\nu\tau}(k_F)_{\mu i\nu i} + (k_F)_{\mu \tau\nu i}(k_F)_{\mu \tau\nu i} + (k_F)_{\mu \tau\nu i}(k_F)_{\mu i\nu\tau}\bigg) \\ 
&= \frac{1}{6}\bigg[\textrm{Tr}(\kappa_{DE}^2)-\textrm{Tr}(\kappa_{DE}\cdot\kappa_{HB})
-3\textrm{Tr}(\kappa_{DB}\cdot\kappa_{HE})-\textrm{Tr}(\kappa_{DE})^2 \bigg],\\
\tilde{A}_3 &= \frac{1}{15} \bigg(  (k_F)_{\mu i\nu i}(k_F)_{\mu j\nu j} +(k_F)_{\mu i\nu j}(k_F)_{\mu i\nu j} + (k_F)_{\mu i\nu j}(k_F)_{\mu j\nu i} \bigg)\\
&= \frac{1}{30}\bigg[\textrm{Tr}(\kappa_{DE}\cdot\kappa_{DE})-\textrm{Tr}(\kappa_{DB}\cdot\kappa_{DB})-4\textrm{Tr}(\kappa_{DB}\cdot\kappa_{HE})
+\frac{7}{2}\textrm{Tr}(\kappa_{HB}\cdot\kappa_{HB})+\textrm{Tr}(\kappa_{DE})^2\bigg].
\end{align}
The integral over $\omega$ can be calculated neglecting vacuum contributions, which arise from the second term in (\ref{2contribution}) and have the same form as they do to first order in $k_F$. The result is given by
\begin{eqnarray}
-\sum_{n,\textbf{p}}\bar{Z}_{LV_2}&=&V\bigg((\tilde{A}_1-\tilde{A}_2+\tilde{A}_3)\frac{\beta^2}{16\pi^2}\bigg(\frac{16\pi^4}{15\beta^5}\bigg)
+(-3\tilde{A}_1+\tilde{A}_2+\tilde{A}_3)\frac{\beta}{8\pi^2}\bigg(\frac{2\pi^4}{15\beta^4}\bigg)\bigg) \\
&=&\bar{K}\bigg(\frac{\pi^2}{45\beta^3}\bigg)V,
\end{eqnarray}
where we have defined
\begin{align}
\nonumber
\bar{K}&\equiv \frac{3}{4}(\tilde{A}_1-3\tilde{A}_2+5\tilde{A}_3)\\
&=\frac{1}{16}\bigg[8\textrm{Tr}(\kappa_{DE})^2-2\textrm{Tr}(\kappa_{DB}\cdot\kappa_{DB})
+10\textrm{Tr}(\kappa_{DB}\cdot\kappa_{HE})- \textrm{Tr}(\kappa_{DE}\cdot\kappa_{DE}) \nonumber \\
&\qquad\quad
+6\textrm{Tr}(\kappa_{DE}\cdot\kappa_{HB})+7\textrm{Tr}(\kappa_{HB}\cdot\kappa_{HB})\bigg] \nonumber \\
&=\frac{1}{4}\bigg[2\textrm{Tr}(\tilde{\kappa}_{o+}\cdot\tilde{\kappa}_{o+})-3\textrm{Tr}(\tilde{\kappa}_{o-}\cdot\tilde{\kappa}_{o-})-\textrm{Tr}(\tilde{\kappa}_{o+}\cdot\tilde{\kappa}_{o-})
+3\textrm{Tr}(\tilde{\kappa}_{e+}\cdot\tilde{\kappa}_{e+})  \nonumber \\
&\qquad\quad
-4\textrm{Tr}(\tilde{\kappa}_{e+}\cdot\tilde{\kappa}_{e-})-4(\tilde{\kappa}_{\textrm{tr}})\textrm{Tr}(\tilde{\kappa}_{o-})+18 (\tilde{\kappa}_{\textrm{tr}})^2 \bigg]
\label{Kfinal}
\end{align}
The partition function of the standard Maxwell theory is well known \cite{Tinfinity0}; neglecting the vacuum contributions it is given by
\begin{equation}
\label{standardplanck}
\textrm{ln Z}_{0} =  -2V \int \frac{d^3p}{(2\pi)^3}\textrm{ln}(1-e^{-\beta\omega})
= \frac{\pi^2}{45\beta^3}V.
\end{equation}
From these results we finally obtain the total partition function
\begin{equation}
\label{partitionfinal}
\textrm{ln Z} = \textrm{ln Z}_{0}+\textrm{ln Z}_{LV}
 = (1+2(k_F)_{\alpha \tau\alpha \tau}+\bar{K})\frac{\pi^2}{45\beta^3}V.
\end{equation}
In order to compare this result with the literature we consider the particular case arising from the isotropic contribution of the parity-even sector, which corresponds to the limit $\textrm{Tr}(\tilde{\kappa}_{o-})=\textrm{Tr}(\tilde{\kappa}_{o+})=\textrm{Tr}(\tilde{\kappa}_{e+})=\textrm{Tr}(\tilde{\kappa}_{e-})=0$, $\tilde{\kappa}_{\textrm{tr}}\neq0$. In this case (\ref{partitionfinal}) reduces to
\begin{equation}
\textrm{ln Z} = \left(1+3(\tilde{\kappa}_{\textrm{tr}})+\frac{9}{2}(\tilde{\kappa}_{\textrm{tr}})^2  \right)\frac{\pi^2}{45\beta^3}V.
\end{equation}
To second order this is the same result reported in \cite{Termeven}.

\bigskip

\begin{figure}[h!]
\centering
	\scalebox{0.8}{\includegraphics{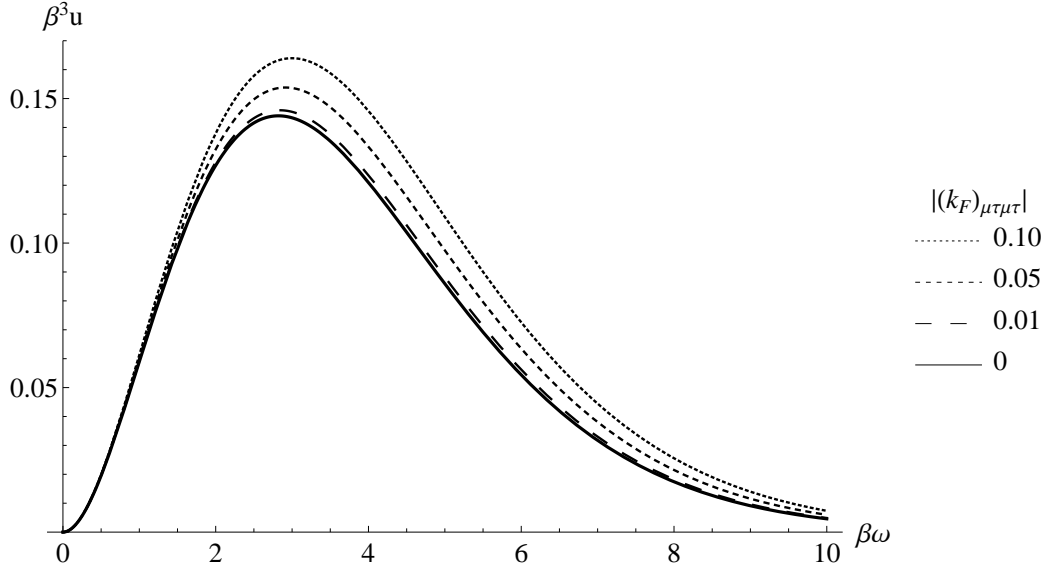}} 
	\caption{Frequency dependence of the photon energy density for non-vanishing CPT-even violating terms, to first order in $k_F$.} 
	\label{bb_fig}
\end{figure}

Starting from (\ref{nonstandardplanck}), (\ref{2contribution}) and (\ref{standardplanck}), the energy density of the photon field can be calculated from the standard thermodynamic relations,
\begin{eqnarray}
u&=&-\frac{1}{V}\frac{\partial \textrm{ln Z}}{\partial \beta} \\ \nonumber
&=& \int_0^\infty d\omega\frac{1}{\pi^2}\frac{\omega^3}{e^{\beta\omega}-1}-\frac{2(k_F)_{\alpha \tau\alpha \tau}}{3\pi^2}\int_0^\infty d\omega \bigg[\frac{\omega^3}{e^{\beta\omega}-1}-\frac{\beta\omega^4 e^{\beta\omega}}{(e^{\beta\omega}-1)^2}\bigg] \\ \nonumber
&&+(\tilde{A}_1-\tilde{A}_2+\tilde{A}_3)\int_0^\infty \frac{d\omega}{16\pi^2}\beta\omega^4\bigg(\beta\omega\, \textrm{coth}\bigg(\frac{\beta\omega}{2}\bigg)-2\bigg)\textrm{csch}^2\bigg(\frac{\beta\omega}{2}\bigg) \\ 
&&-(-3\tilde{A}_1+\tilde{A}_2+\tilde{A}_3) \int_0^\infty \frac{d\omega}{4\pi^2} \bigg[\frac{\omega^3}{e^{\beta\omega}-1}-\frac{\beta\omega^4 e^{\beta\omega}}{(e^{\beta\omega}-1)^2}\bigg].
\end{eqnarray}
This implies a modification to the Planck distribution, where now the frequency dependence of the energy density is given by
\begin{eqnarray}
\label{Planck}
u(\omega)&=&\frac{1}{\pi^2}\frac{\omega^3}{e^{\beta\omega}-1} 
-\frac{1}{\pi^2}\bigg(\frac{2}{3}(k_F)_{\alpha \tau\alpha \tau} +\frac{1}{4}(-3\tilde{A}_1+\tilde{A}_2+\tilde{A}_3) \bigg)\bigg[\frac{\omega^3}{e^{\beta\omega}-1}-\frac{\beta\omega^4 e^{\beta\omega}}{(e^{\beta\omega}-1)^2}\bigg] \\ \nonumber &&+ \frac{1}{16\pi^2}(\tilde{A}_1-\tilde{A}_2+\tilde{A}_3)\beta\omega^4\bigg(\beta\omega\, \textrm{coth}\bigg(\frac{\beta\omega}{2}\bigg)-2\bigg)\textrm{csch}^2\bigg(\frac{\beta\omega}{2}\bigg),
\end{eqnarray}
(see Fig. \ref{bb_fig}). The total energy density is obtained integrating the last equation or deriving directly (\ref{partitionfinal})
\begin{equation}
u(T)=(1+2(k_F)_{\alpha \tau\alpha \tau}+\bar{K})\frac{\pi^2}{15}T^4\, .
\end{equation}
The thermodynamic relations
\begin{equation}
U=-\frac{\partial }{\partial \beta}\textrm{ln Z},\quad\quad\quad P=T\frac{\partial }{\partial V}\textrm{ln Z},
\end{equation}
together with (\ref{partitionfinal}), imply that the equation of state remains unchanged,
\begin{equation}
PV=\frac{U}{3}\,.
\end{equation}

\bigskip

The functional scheme that we used to derive the photon energy density is not the only prescription available. It is possible to reproduce the previous results following a classical thermodynamic scheme, as we now show. Since this computation is not the main purpose of our study, we only calculate the energy density to first order in $k_F$. We follow a similar treatment as \cite{ModosPlanck}. By adopting a thermal distribution for the photons and dispersion law (\ref{dispersion}), the spectral energy density per frequency (and per polarization)  in the semiclassical phase space is given by
\begin{equation}
u(\omega_\pm)d\omega_\pm=\frac{1}{(2\pi)^3}\frac{\hbar\omega_\pm}{e^{\hbar\omega_\pm/k_BT}-1}k^2dk\ d\Omega\, .
\end{equation}
From (\ref{dispersion}), we have $\omega_\pm=(1+\delta_\pm)|\vec{p}|$, where $\delta_\pm=\rho\pm\sigma$. For the isotropic frequencies, one can immediately substitute $\int d\Omega \rightarrow 4\pi$, however we now have an angular dependence for the frequency, $\omega_\pm=\omega_\pm(\theta,\phi)$. To first order in $\delta_\pm\ll1$ and using $|k|=\omega$, we find
\begin{equation}
u(\omega_\pm)\,d\omega_\pm=\frac{1}{(2\pi)^3}\bigg(\frac{\hbar\omega^3}{e^{\hbar\omega/k_BT}-1}\bigg)d\omega\, d\Omega
+\frac{1}{(2\pi)^3}\bigg(\frac{\hbar\omega^3}{e^{\hbar\omega/k_BT}-1}-\frac{\hbar^2}{k_BT}\frac{\omega^4 e^{\hbar\omega/k_BT}}{(e^{\hbar\omega/k_BT}-1)^2}\bigg)\delta_\pm d\omega\, d\Omega\, .
\end{equation}
If we sum over both modes under the assumption that each one contributes equally, we have that the Lorentz violation contribution to the energy density is given by
\begin{equation}
[u(\omega_+)d\omega_+]_{LV}+[u(\omega_-)d\omega_-]_{LV}=
\frac{1}{(2\pi)^3}\bigg(\frac{\hbar\omega^3}{e^{\hbar\omega/k_BT}-1}-\frac{\hbar^2}{k_BT}\frac{\omega^4 e^{\hbar\omega/k_BT}}{(e^{\hbar\omega/k_BT}-1)^2}\bigg)(\delta_++\delta_-) d\omega\, d\Omega\,.
\label{Energyviol}
\end{equation}
Since $\delta_++\delta_-=2\rho=-\tilde{k}_\alpha\,^\alpha$ and
\begin{eqnarray}
\tilde{k}_\alpha\,^\alpha\quad \xrightarrow{\quad-i\tau\quad}  \quad (k_F)_{\alpha\tau \alpha\tau}-2i(k_F)_{j\tau j i}p_i-(k_F)_{\alpha i \alpha j}\hat{p}_i\hat{p}_j,
\end{eqnarray}
where $\alpha=\tau,1,2,3$, upon integration of (\ref{Energyviol}) over the solid angle $d\Omega$ we finally have
\begin{eqnarray}
[u(\omega)d\omega]_{LV}&=&[u(\omega_+)d\omega_+]_{LV}+[u(\omega_-)d\omega_-]_{LV} \\ \nonumber &=&
-\frac{4\pi}{(2\pi)^3}((k_F)_{\alpha\tau \alpha\tau}-\frac{1}{3}(k_F)_{\alpha i \alpha i})\bigg(\frac{\hbar\omega^3}{e^{\hbar\omega/k_BT}-1}-\frac{\hbar^2}{k_BT}\frac{\omega^4 e^{\hbar\omega/k_BT}}{(e^{\hbar\omega/k_BT}-1)^2}\bigg) d\omega \\ \nonumber
&=& -\frac{2}{3\pi^2}((k_F)_{\alpha\tau \alpha\tau})\bigg(\frac{\hbar\omega^3}{e^{\hbar\omega/k_BT}-1}-\frac{\hbar^2}{k_BT}\frac{\omega^4 e^{\hbar\omega/k_BT}}{(e^{\hbar\omega/k_BT}-1)^2}\bigg) d\omega, \\ \nonumber
\end{eqnarray}
where in the second line we have used the condition of vanishing double trace in the imaginary time (\ref{vanishingtrace}). Taking $\hbar=1$ and $\beta=1/k_BT$, we find a result in agreement with (\ref{Planck}).

\section{Summary}\label{summary}

We have presented in this paper a study of the Hamiltonian formalism and the canonical quantization, via the Dirac formalism, of the CPT-even photon sector of the Lorentz-violating Standard Model Extension. We have found that the gauge freedom of standard electrodynamics is not lost in the CPT-even photon sector of the SME. Additionally, in analogy with the standard electromagnetic case, this model possesses first class constraints which restrict the size of phase space; however, the form of these constraints is different from the standard case when they  are written in terms of the electric and magnetic fields. The gauge fixing procedure in the CPT-even photon sector of the SME does not significantly differ from the standard electrodynamics case.

The partition function was explicitly calculated to first  and second order in the Lorentz-violating parameters of the CPT-even photon sector of the SME, for an arbitrary configuration of the coefficients $(k_F)_{\mu\nu\alpha\beta}$; we have presented our results in terms of the defining $(k_F)_{\mu\nu\alpha\beta}$ coefficients as well as the two parametrizations (\ref{parametrization1}) and (\ref{parametrization21}-\ref{parametrization22}). Making use of the standard thermodynamic relations we found corrections to the blackbody radiation law. These include corrections with the same functional dependence on frequency, as well as corrections with a different functional dependence. The total temperature-dependent energy density receives a correction with the same temperature dependence as the standard case, but we found that the equation of state has the same form of the standard electrodynamic theory. This corresponds to a cosmological thermal history consistent with the standard scenario. The result was verified employing a classical thermodynamic scheme. We have obtained that all modifications arising from Lorentz violation to first order are proportional to $(k_F)_{\alpha 0 \alpha 0}$ coefficients, there is no contribution to first order from the $(k_F)_{0 i \alpha\beta}$ sector to the partition function as well as the density energy.  We have compared our general results with a particular case previously obtained in \cite{Termeven} for the isotropic contribution of the parity-even sector, which is encoded into the coefficient $(k_F)_{i 0 i 0}\approx\tilde{\kappa}_{\textrm{tr}}$, finding a complete agreement between both works. Furthermore, following the scheme employed here, it is possible to obtain the contributions arising from the couplings of the distinct sectors of the full CPT-even photon sector of the SME, as explicitly shown in Eq.~(\ref{Kfinal}).

\section{Acknowledgements}

C.A. Escobar acknowledges support from a CONACyT graduate fellowship and  thanks L.F Urrutia for many valuable discussions, comments and suggestions.

\end{document}